\documentclass[%
 preprint,
 amsmath,amssymb,
 aps,
]{revtex4-2}

\usepackage{graphicx}
\usepackage{dcolumn}
\usepackage{bm}

\usepackage{color}

\begin{document}

\preprint{APS/123-QED}

\title{Theoretical predictions of melting behaviors of hcp iron up to 4000 GPa}

\author{Tran Dinh Cuong}
\email{cuong.trandinh@phenikaa-uni.edu.vn}
\affiliation{Faculty of Materials Science and Engineering, Phenikaa University, Hanoi 12116, Vietnam.}
\author{Nguyen Quang Hoc}%
\affiliation{Faculty of Physics, Hanoi National University of Education, Hanoi 11310, Vietnam.}%
\author{Nguyen Duc Trung}%
\affiliation{Faculty of Physics, Hanoi National University of Education, Hanoi 11310, Vietnam.}%
\author{Nguyen Thi Thao}%
\affiliation{Faculty of Physics, Hanoi National University of Education, Hanoi 11310, Vietnam.}%
\author{Anh D. Phan}
\affiliation{Faculty of Materials Science and Engineering, Phenikaa University, Hanoi 12116, Vietnam.}%
\affiliation{Phenikaa Institute for Advanced Study (PIAS), Phenikaa University, Hanoi 12116, Vietnam.}%

\date{\today}

\begin{abstract}
The high-pressure melting diagram of iron is a vital ingredient for the geodynamic modeling of planetary interiors. Nonetheless, available data for molten iron show an alarming discrepancy. Herein, we propose an efficient one-phase approach to capture the solid-liquid transition of iron under extreme conditions. Our basic idea is to extend the statistical moment method to determine the density of iron in the TPa region. On that basis, we adapt the work-heat equivalence principle to appropriately link equation-of-state parameters with melting properties. This strategy allows explaining cutting-edge experimental and \textit{ab initio} results without massive computational workloads. Our theoretical calculations would be helpful to constrain the chemical composition, internal dynamics, and thermal evolution of the Earth and super-Earths.
\end{abstract}

\maketitle


\section{Introduction}
The melting process of compressed iron has acquired immense attention from the geophysical community. According to seismological and cosmo-chemical evidence, more than 80 \% of the Earth's core is constituted of this transition metal \cite{1}. Hence, its melting information between 135 and 330 GPa is crucial for considering the Earth's thermal profile \cite{2}, heat flow \cite{3}, and magnetic field \cite{4}. Furthermore, in recent years, 1569 super-Earths have been discovered outside our solar system \cite{5}. Since these exoplanets can be tenfold heavier than the Earth, investigating their formation, evolution, and habitability requires a deep understanding of molten iron at extraordinary pressures up to 4000 GPa \cite{6}.

The past four decades have witnessed arduous efforts to explore the melting characteristics of iron in deep-planetary interiors. On the experimental side, diamond anvil cell (DAC) \cite{7,8,9} and shock-wave (SW) \cite{10,11,12,13,14} techniques have been continuously improved. Unfortunately, the accurate melting curve of iron is clouded by numerous controversial results \cite{15}. At 330 GPa, the reported melting temperatures exhibit a drastic scatter from 4850 to 7600 K. Therefore, determining the Earth's structure, dynamics, and history remains a major problem. Besides, it is challenging to access the TPa regime via current DAC and SW measurements \cite{16,17,18}. On the simulation side, the density functional theory (DFT) and the diffusion Monte Carlo (DMC) method have been developed to yield insights into the melting phenomenon \cite{19,20,21}. These powerful \textit{ab initio} approaches can evaluate both ionic and electronic contributions without empirical adjustable parameters. Nevertheless, using DFT and DMC codes necessitates enormous computational resources \cite{22}. In addition, \textit{ab initio} phase relations near the Earth's inner core boundary (ICB) are still hotly debated \cite{23,24,25}.  

Lately, the statistical moment method (SMM) has emerged as a promising quantum-mechanical tool for predicting the melting behaviors of metallic systems \cite{26,27,28}. The SMM can describe the atomic rearrangement under various thermodynamic conditions via the quantum density matrix \cite{29,30,31}. Based on the SMM equation of state (EOS), one often locates the melting point by analyzing the instability of the solid phase \cite{26,27,28}. These approximations help enhance the computational efficiency dramatically \cite{32}. A complete SMM run only takes ten seconds on a typical computer. However, SMM outputs for iron are inconsistent with each other \cite{33,34,35,36}. Even at 135 GPa, the difference can be up to 30 \%. Moreover, existing SMM analyses \cite{33,34,35,36} are limited to the ICB pressure of 330 GPa. Consequently, the melting mechanism of iron inside super-Earths stays a mystery to SMM researchers. 

Herein, we formulate a simple but effective scheme to overcome SMM shortcomings. Theoretical calculations are primarily carried out for hcp iron because it is the leading candidate for planetary cores \cite{37,38,39,40}. Our study is thoroughly compared with prior DAC, SW, \textit{ab initio}, and SMM works. 

\section{Equation of state}
According to Cuong and Phan \cite{35}, the cohesive energy $E_i$ of the $i$-th iron atom can be well parameterized by the Morse model \cite{41,42,43} as
\begin{eqnarray}
E_i=\frac{1}{2}\sum_{j\ne i}D\left[e^{-2\alpha\left(r_{ij}-r_0\right)}-2e^{-\alpha\left(r_{ij}-r_0\right)}\right],
\label{eq:1}
\end{eqnarray}
where $D$ is the potential depth, $\alpha^{-1}$ is the potential width, and $r_0$ is the critical value of the interatomic separation $r_{ij}$. This treatment is useful for accelerating computational processes and elucidating  vibrational properties \cite{35}. Thus, we keep utilizing the Morse function \cite{41,42,43} to construct the SMM EOS for hcp iron. Specifically, $D=0.6317$ eV, $\alpha=1.4107$ \AA$^{-1}$, and $r_0=2.6141$ \AA\ are extracted from recent DFT simulations \cite{44}. On that basis, it is feasible to deduce quasi-harmonic and anharmonic force constants from the Leibfried-Ludwig theory \cite{45} as
\begin{eqnarray}
k_i=\left(\frac{\partial^2E_i}{\partial u_{i\delta}^2}\right)_{eq},\quad\beta_i=\left(\frac{\partial^3E_i}{\partial u_{i\delta}\partial u_{i\eta}^2}\right)_{eq},\nonumber\\
\gamma_{1i}=\left(\frac{\partial^4E_i}{\partial u_{i\delta}^4}\right)_{eq},\quad\gamma_{2i}=\left(\frac{\partial^4E_i}{\partial u_{i\delta}^2u_{i\eta}^2}\right)_{eq},
\label{eq:2}
\end{eqnarray}
where $u_{i\delta}$ and $u_{i\eta}$ indicate atomic displacements along Cartesian axes ($\delta\ne\eta=x,y,z$). For simplicity, their average values are supposed to satisfy the following symmetric criterion
\begin{eqnarray}
\langle u_{i\delta}\rangle=\langle u_{i\eta}\rangle=\langle u_{i}\rangle .
\label{eq:3}
\end{eqnarray}

In the thermodynamic equilibrium, if the $i$-th atom is impacted by a supplemental force $f_i$, its movement will be governed by \cite{46}
\begin{eqnarray}
k_i\langle u_{i}\rangle+\beta_i\langle u_{i}^2\rangle+\left(\frac{1}{6}\gamma_{1i}+\gamma_{2i}\right)\langle u_{i}^3\rangle-f_i=0.
\label{eq:4}
\end{eqnarray}
Interestingly, the SMM enables us to associate the first-order moment ($\langle u_{i}\rangle$) with its high-order counterparts ($\langle u_{i}^2\rangle$ and $\langle u_{i}^3\rangle$) via \cite{29,30,31}
\begin{eqnarray}
\langle u_{i}^2\rangle=\langle u_{i}\rangle^2+\theta\frac{\partial\langle u_{i}\rangle}{\partial f_i}+\frac{\theta}{k_i}\left(x_i\coth x_i-1\right),
\label{eq:5}
\end{eqnarray}
\begin{eqnarray}
\langle u_{i}^3\rangle=\langle u_{i}\rangle^3+3\theta\langle u_{i}\rangle\frac{\partial\langle u_{i}\rangle}{\partial f_i}+\theta^2\frac{\partial^2\langle u_{i}\rangle}{\partial f_i^2}+\frac{\theta}{k_i}\left(x_i\coth x_i-1\right)\langle u_{i}\rangle,
\label{eq:6}
\end{eqnarray}
where $\theta=k_BT$ is the Boltzmann constant $k_B$ times the absolute temperature $T$, $x_i=\hbar\omega_{i}/2\theta$ denotes the scaled phonon energy, $\hbar$ stands for the reduced Planck constant, and $\omega_i$ represents the Einstein frequency. Inserting these statistical correlations into Equation (\ref{eq:4}) gives us
\begin{gather}
\gamma_i\theta^2\frac{d^2y_i}{df_i^{*2}}+3\gamma_i\theta y_i\frac{dy_i}{df_i^{*}}+\gamma_iy_i^3+K_iy_i+\frac{\gamma_i\theta}{k_i}\left(x_i\coth x_i-1\right)y_i-f_i^*=0,
\label{eq:7}
\end{gather}
where
\begin{gather}
\gamma_i=\frac{1}{6}\gamma_{1i}+\gamma_{2i},\quad K_i=k_i-\frac{\beta_i^2}{3\gamma_i},\quad y_i=\langle u_{i}\rangle+\frac{\beta_i}{3\gamma_i},\nonumber\\
f_i^*=f_i-\frac{\beta_ik_i}{3\gamma_i}\left[\frac{2\beta_i^2}{9\gamma_ik_i}+\frac{2\gamma_i\theta}{k_i^2}\left(x_i\coth x_i-1\right)-1\right].
\label{eq:8}
\end{gather}
Notably, since $f_i^*$ is arbitrarily small, we can express $y_i$ as a polynomial function of $f_i^*$, which is
\begin{gather}
y_i=y_{i0}+A_{1i}f_i^*+A_{2i}f_i^{*2},
\label{eq:9}
\end{gather}
where $A_{1i}$ and $A_{2i}$ are Taylor-Maclaurin coefficients. Applying the Tang-Hung iterative technique \cite{47} to Equations (\ref{eq:7})-(\ref{eq:9}) provides
\begin{gather}
y_{i0}=\sqrt{\frac{2\gamma_i\theta^2}{3K_i^3}A_i},\nonumber\\
A_{1i}=\frac{1}{K_i}\left[1+\frac{2\gamma_i^2\theta^2}{K_i^4}\left(1+\frac{x_i\coth x_i}{2}\right)\left(1+x_i\coth x_i\right)\right],\nonumber\\
A_{2i}=\frac{1}{2K_iy_{i0}}\left[\frac{1}{3K_i}\left(1-x_i\coth x_i\right)-\frac{1}{K_i}-\frac{y_{i0}^2}{\theta}\right].
\label{eq:10}
\end{gather}
The explicit form of $A_i$ was extensively reported in previous SMM literature \cite{48,49,50}. Consequently, when $f_i$ goes to zero, $\langle u_{i}\rangle$ is estimated by
\begin{gather}
\langle u_{i}\rangle\approx\sqrt{\frac{2\gamma_i\theta^2}{3K_i^3}A_i}-\frac{\beta_i}{3\gamma_i}+A_{1i}\frac{\beta_ik_i}{3\gamma_i}\left[1-\frac{2\beta_i^2}{9\gamma_ik_i}-\frac{2\gamma_i\theta}{k_i^2}\left(x_i\coth x_i-1\right)\right].
\label{eq:11}
\end{gather}
In other words, we can obtain the density $\rho$ of hcp iron from \cite{35}
\begin{gather}
\rho(P,T)=\frac{m_i\sqrt{2}}{\left[a_i(P,0)+\langle u_{i}\rangle(P,T)\right]^3},
\label{eq:12}
\end{gather}
where $P$ is the hydrostatic pressure, $m_i$ is the atomic mass, and $a_i$ is the mean nearest-neighbor distance.

\begin{figure}[htp]
\includegraphics[width=12 cm]{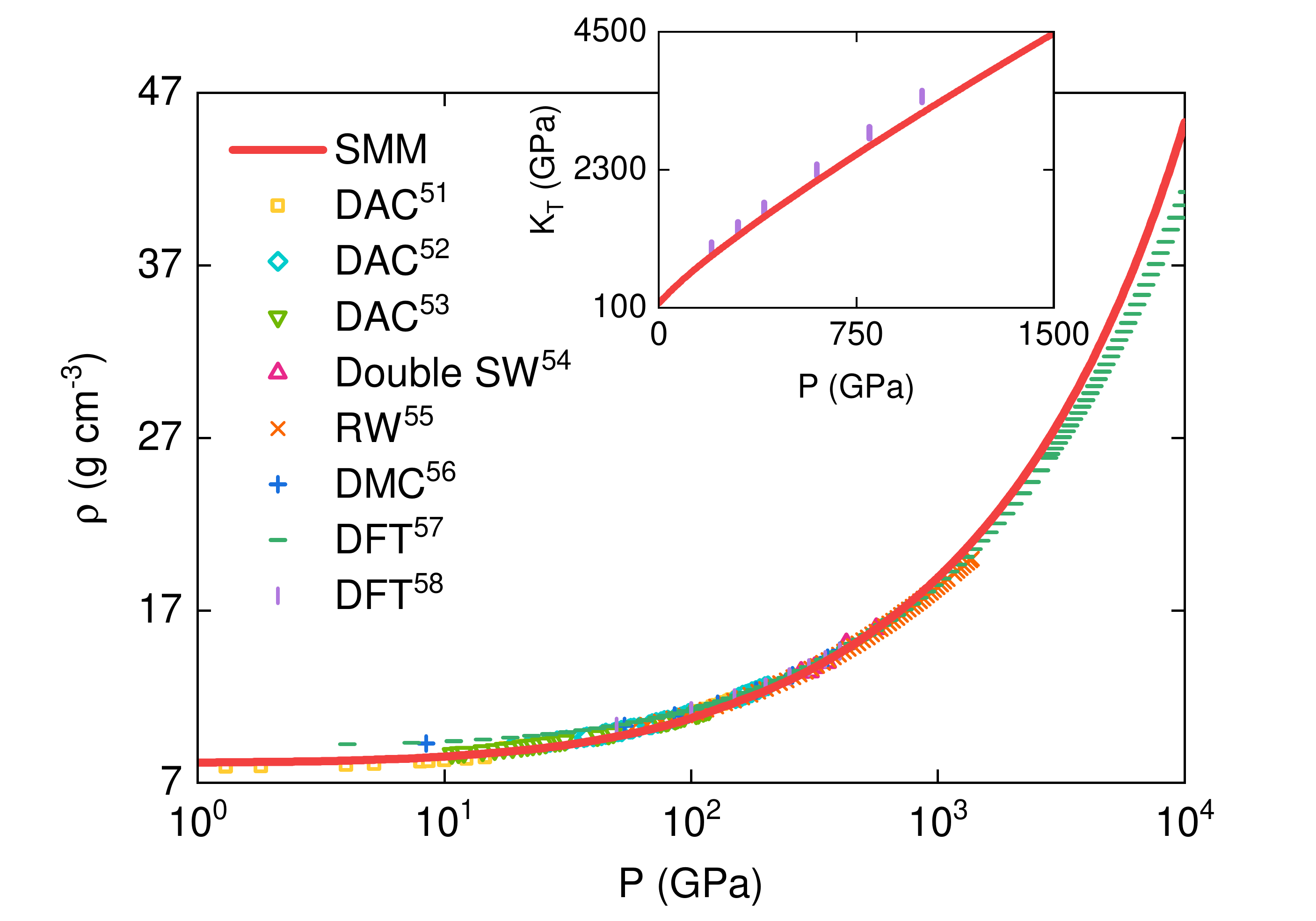}
\caption{\label{fig:1}(Color online) The room-temperature SMM EOS of iron in comparison with recent measured and simulated results \cite{51,52,53,54,55,56,57,58}. Inset: The pressure dependence of the isothermal bulk modulus given by SMM analyses and DFT calculations \cite{58}.}
\end{figure}

Numerical calculations relying on Equation (\ref{eq:12}) are shown in Figure \ref{fig:1}. It is conspicuous that hcp iron becomes significantly denser during isothermal squeezing. At room temperature, a sharp growth of 37.17 $\mathrm{g}\,\mathrm{cm}^{-3}$ is recorded in $\rho$ between 0 and 10000 GPa. Our SMM analyses agree quantitatively with static DAC measurements \cite{51,52,53}, multi-shock experiments \cite{54}, ramp-wave (RW) investigations \cite{55}, and DFT/DMC computations \cite{56,57,58}. The deviation between the SMM and other state-of-the-art methods \cite{51,52,53,54,55,56,57,58} is less than 10 \% over a vast pressure range. Better consensus can be reached by adding magnetic terms \cite{51,59} to \textit{ab initio} EOSs \cite{56,57,58} at $P\leq50$ GPa. Hence, our SMM results would be valuable for constraining the chemical composition of extrasolar bodies up to ten Earth masses (see Appendix A). To facilitate geodynamic modeling, we fit SMM data by the Vinet EOS \cite{60} as
\begin{gather}
P=3K_0\xi^{-2}(1-\xi)\exp\left[\frac{3}{2}\left(K_0'-1\right)\left(1-\xi\right)\right],
\label{eq:13}
\end{gather}
where $\xi=\left(\rho_0/\rho\right)^{1/3}\,$, $\rho_0=8.13$ $\mathrm{g}\,\mathrm{cm}^{-3}$, $K_0=175.34$ GPa, and $K_0'=5.02$. Equation (\ref{eq:13}) allows us to consider compression effects on the bulk modulus $K_T$ by
\begin{gather}
K_T=K_{0}\left[\frac{3\left(1-K_0'\right)}{2}+\frac{3K_0'-5}{2\xi}+\frac{2}{\xi^2}\right]\exp\left[\frac{3}{2}\left(K_0'-1\right)\left(1-\xi\right)\right].
\label{eq:14}
\end{gather}
Excitingly, even for $K_T$, the SMM output only differs from the latest DFT one \cite{58} by 7.58 \% at 1000 GPa. This consistency reaffirms the reliability of the SMM EOS. 

\section{Melting behaviors}
It cannot be denied that there is a tight connection between the EOS and other physical properties. In particular, numerous attempts have been made to derive melting quantities from EOS parameters \cite{61,62,63}. Lately, Ma \textit{et al.} \cite{64} have introduced a semi-empirical approach, called the work-heat equivalence principle (WHEP), to handle this long-standing issue. In the WHEP picture \cite{64}, a liquid can be crystallized owing to exothermic or distortion influences. Therefore, heat energies during isobaric cooling are analogous to mechanical works during isothermal squeezing. This unique point of view \cite{64} helps determine the melting temperature $T^m$ by
\begin{gather}
T^m(P)=T_{1}\left(P_1\right)+\left[T_2\left(P_2\right)-T_1\left(P_1\right)\right]\sqrt{\cfrac{\displaystyle\int_{\xi_1}^{\xi}P\left(\xi'\right)\xi'^2d\xi'}{\displaystyle\int_{\xi_1}^{\xi_2}P\left(\xi'\right)\xi'^2d\xi'}},
\label{eq:15}
\end{gather}
where $\left(T_1,P_1,\xi_1\right)$ and $\left(T_2,P_2,\xi_2\right)$ are two reference melting points. Based on the Murnaghan approximation \cite{65}, Ma \textit{et al.} \cite{64} have explicitly rewritten Equation (\ref{eq:15}) by 
\begin{gather}
T^m(P)=T_{1}\left(P_1\right)+\left[T_2\left(P_2\right)-T_1\left(P_1\right)\right]\sqrt{\cfrac{\left(\cfrac{K_0'P+K_0}{K_0'P_1+K_0}\right)^{-\frac{1}{K_0'}}\left(\cfrac{P+K_0}{P_1+K_0}\right)-1}{\left(\cfrac{K_0'P_2+K_0}{K_0'P_1+K_0}\right)^{-\frac{1}{K_0'}}\left(\cfrac{P_2+K_0}{P_1+K_0}\right)-1}}.
\label{eq:16}
\end{gather}
Equation (\ref{eq:16}) has been successfully applied to various metallic substances \cite{64} and ionic compounds \cite{66}. Notwithstanding, all WHEP predictions \cite{64,66} have been restricted to $P\sim2K_0$. A viable solution for the WHEP problem is to replace the Murnaghan EOS with the SMM EOS \cite{63}. Combining Equation (\ref{eq:13}) and Equation (\ref{eq:15}) yields
\begin{gather}
T^m(P)=T_{1}\left(P_1\right)+\left[T_2\left(P_2\right)-T_1\left(P_1\right)\right]\sqrt{\cfrac{W\left(\xi\right)-W\left(\xi_1\right)}{W\left(\xi_2\right)-W\left(\xi_1\right)}},\nonumber\\
W\left(\xi\right)=\left[1+\frac{3}{2}\left(K_0'-1\right)\left(\xi-1\right)\right]\exp\left[\frac{3}{2}\left(K_0'-1\right)\left(1-\xi\right)\right].
\label{eq:17}
\end{gather}
The effectiveness of Equation (\ref{eq:17}) is thoroughly demonstrated in the Supplemental Material \cite{67}. In this section, Equation (\ref{eq:17}) is employed to decode the melting behaviors of iron at deep-planetary conditions. For objectivity, we choose $P_1=0$ GPa, $T_1=1811$ K, $P_2=5.2$ GPa, and $T_2=1991$ K \cite{68}. This low-pressure melting information has been widely validated by different research groups \cite{69,70}. 

\begin{figure}[htp]
\includegraphics[width=12 cm]{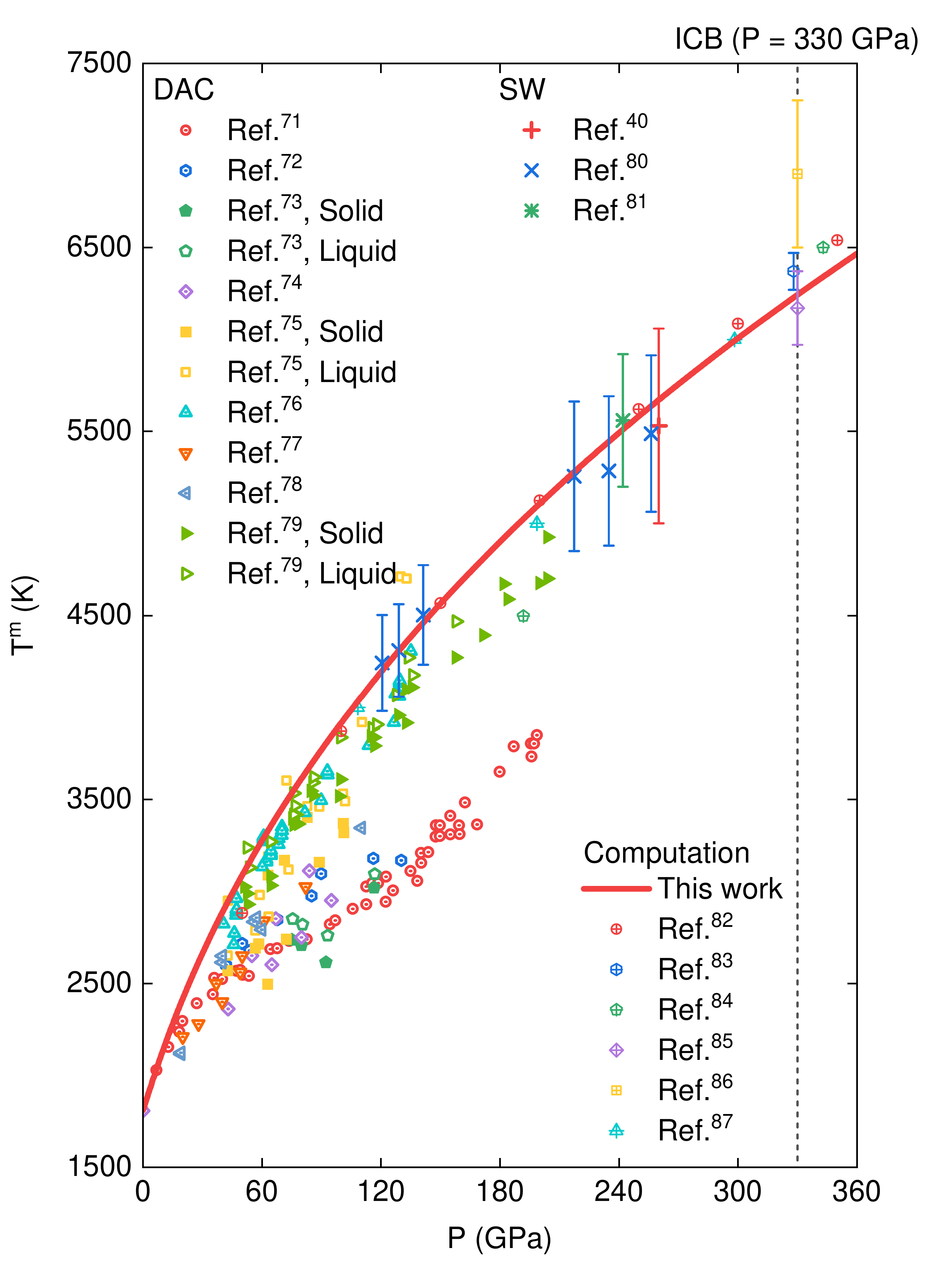}
\caption{\label{fig:2}(Color online) The melting profile of iron inferred from our SMM-WHEP analyses, DAC experiments \cite{71,72,73,74,75,76,77,78,79}, SW measurements \cite{40,80,81}, \textit{ab initio} calculations \cite{82,83,84,85,86}, and atomistic simulations adopting machine-learning potentials \cite{87} in a pressure range from 0 to 360 GPa.}
\end{figure}

Figure \ref{fig:2} shows the melting properties of iron inside the Earth. In general, there are three principal scenarios for its melting process under extreme compression. Firstly, Boehler \textit{et al.} \cite{71,72}, Aquilanti \textit{et al.} \cite{73}, and Basu \textit{et al.} \cite{74} have recommended a relatively flat melting curve for iron. They have predicted a slight increase in $T^m$ to $4850\pm200$ K at 330 GPa \cite{71,72,73,74}. However, their DAC data \cite{71,72,73,74} may have been misinterpreted due to the contamination \cite{75} or deformation \cite{76} of iron samples. Secondly, Jackson \textit{et al.} \cite{77} and Zhang \textit{et al.} \cite{78} have suggested that iron should melt in the intermediate-temperature region. Extrapolating their DAC results \cite{77,78} to 330 GPa has provided $T^m=5700\pm200$ K. Unfortunately, this value may have been underestimated because of the inaccurate treatment of thermal pressures \cite{75}. Finally, a steep melting line for iron has been explored by Anzellini \textit{et al.} \cite{79}, Morard \textit{et al.} \cite{75}, and Hou \textit{et al.} \cite{76}. Their DAC studies \cite{75,76,79} have found $T^m=6230\pm500$ K at the ICB. This melting characteristic of iron has been reconfirmed by cutting-edge SW measurements of Li \textit{et al.} \cite{80} ($T^m=5950\pm400$), Turneaure \textit{et al.} \cite{81} ($T^m\sim6400$ K), and Kraus \textit{et al.} \cite{40} ($T^m=6230\pm540$ K). 

Our SMM-WHEP calculations strongly support the last experimental scenario \cite{40,75,76,79,80,81}. Specifically, the SMM-WHEP melting temperature climbs sharply from 4384 K at 135 GPa to 6243 K at 330 GPa. This melting tendency is in line with our two-phase simulations in the Supplemental Material \cite{67}. Furthermore, excellent accordance among SMM-WHEP, DFT \cite{82,83,84,85}, DMC \cite{86}, and machine-learning \cite{87} outputs is achieved. Thus, the true melting profile of iron may be extremely close to our solid-liquid boundary. Our SMM-WHEP findings would have profound geophysical implications for the Earth's dynamics and evolution (see Appendix B).

\begin{figure}[htp]
\includegraphics[width=12 cm]{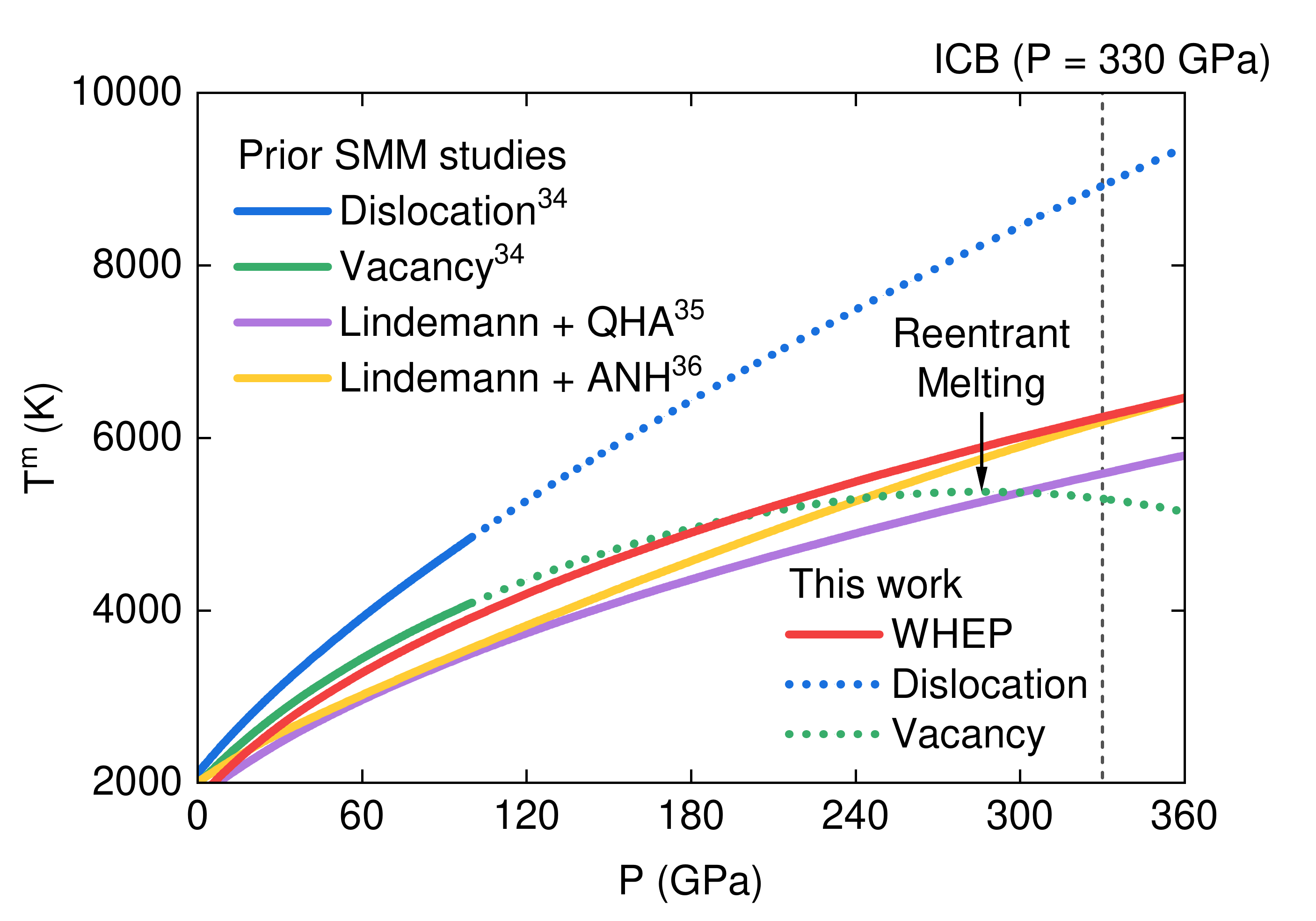}
\caption{\label{fig:3}(Color online) Influences of hydrostatic compression on the melting point of iron derived from our current research and previous SMM investigations \cite{34,35,36}. QHA and ANH stand for quasi-harmonic and anharmonic approximations, respectively.}
\end{figure}

Remarkably, this is the first time the SMM and other approaches \cite{40,75,76,79,80,81,82,83,84,85,86,87} have been consolidated in a unified picture. For clarity, we compare available SMM melting diagrams in Figure \ref{fig:3}. Pioneering SMM works \cite{33,34} have utilized the dislocation-mediated melting theory \cite{63,88,89} to investigate the melting event of iron as
\begin{gather}
\frac{\mu}{\rho T^m}=\mathrm{const},
\label{eq:18}
\end{gather}
where $\mu$ is the shear modulus. Although Equation (\ref{eq:18}) can be straightforwardly solved, the obtained melting point is often overestimated \cite{90,91,92}. For example, at the ICB, Equation (\ref{eq:18}) gives us $T^m=8923$ K, which is far beyond \textit{ab initio} outputs \cite{82,83,84,85,86}. To address the mentioned drawback, Hoc \textit{et al.} \cite{34} have taken into account the impact of point defects by 
\begin{gather}
\Delta T^m=\left(\frac{\partial T}{\partial c_V}\right)_{\rho,P}c_V=\frac{2k_BT^{m2}}{T^m\cfrac{\partial E_i}{\partial T}-E_i},
\label{eq:19}
\end{gather}
where $\Delta T^m$ is the temperature correction and $c_V$ is the monovacancy concentration. The SMM analyses of Hoc \textit{et al.} \cite{34} are compatible with the DFT free-energy computations of Alfe \textit{et al.} \cite{82} at $P\leq100$ GPa. Nonetheless, when extending their idea to denser systems, we observe the presence of reentrant melting at 286 GPa and 5376 K. This strange phenomenon is most likely a consequence of ignoring the vacancy formation volume \cite{28}. 

Apart from defective models \cite{33,34}, the melting transition of iron has been related to its vibrational instability via the Lindemann criterion \cite{93,94,95} as
\begin{gather}
\frac{\langle u_{i}^2\rangle}{a_i^2}=\mathrm{const}.
\label{eq:20}
\end{gather}
Cuong and Phan \cite{35} have accurately reproduced high-quality DAC data of Morard \textit{et al.} \cite{75} by solving Equation (\ref{eq:20}) within the quasi-harmonic approximation. Nevertheless, a slight underestimation of the melting temperature has been recorded at $P\geq135$ GPa \cite{35}. Tan and Tam \cite{36} have revealed that the situation can be improved by performing fully anharmonic SMM calculations. As presented in Figure 3, their numerical results \cite{36} are quantitatively consistent with our SMM-WHEP predictions. However, from our perspective, the above agreement is just an accident. Indeed, Tan and Tam \cite{36} have used the Lennard-Jones pairwise potential \cite{96} to formulate the SMM EOS. This method often causes an overestimation of the scaled atomic volume \cite{97,98,99}. At 330 GPa, $\xi^3$ for the Lennard-Jones iron has been reported to be 0.72 \cite{36}, about 22 \% larger than the DAC value \cite{51,52,53}.

Figure \ref{fig:4} illustrates the melting curve of iron under super-Earth conditions. Overall, SMM-WHEP outputs are somewhat lower than their DFT counterparts \cite{84}. At 1517 GPa, our present SMM-WHEP analyses provide $T^m=11560$ K, whereas earlier DFT simulations of Bouchet \textit{et al.} \cite{84} have yielded $T^m=12506$ K. This discrepancy primarily originates from the selection of initial crystalline structures. Unlike us, Bouchet \textit{et al.} \cite{84} have focused on bcc iron. According to Belonoshko \textit{et al.} \cite{25}, the bcc-liquid boundary can be slightly higher than the hcp-liquid one. To be more specific, we reapply the SMM-WHEP scheme to the bcc phase with $0\leq P\leq4000$ GPa. Fascinatingly, while $K_0'$ is almost unchanged, $K_0$ rises by a factor of 1.13. These findings are bolstered by modern thermodynamic calculations of Dorogokupets \textit{et al.} \cite{70}. Hence, our high-pressure melting line becomes steeper and passes through all DFT points \cite{84} within the error bars. 

\begin{figure}[htp]
\includegraphics[width=12 cm]{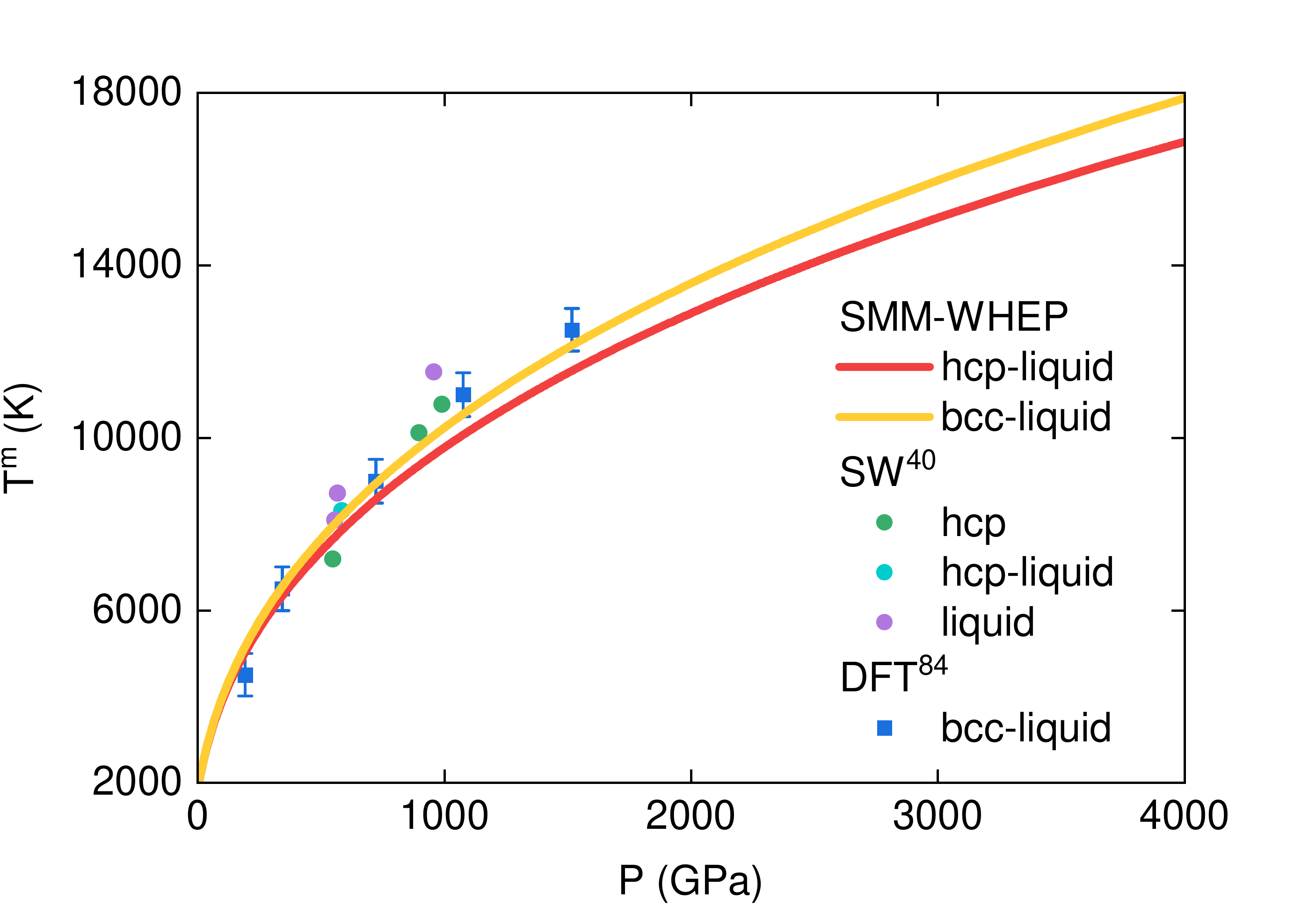}
\caption{\label{fig:4}(Color online) The melting temperature of iron as a function of super-Earth pressure gained from our SMM-WHEP calculations, laser-driven SW experiments \cite{40}, and DFT two-phase simulations \cite{84}.}
\end{figure}

Despite the achieved agreement, it should be emphasized that the real phase diagram of iron remains mysterious. Experimentalists have only detected the melting signatures of hcp iron at above 100 GPa \cite{40}. On the other hand, large-scale \textit{ab initio} computations have indicated that bcc iron can be dynamically stabilized due to superionic-like diffusion at elevated temperatures \cite{23}. More efforts are essential to answer what phase of iron exists in planetary cores.  

For convenience, we parameterize SMM-WHEP melting profiles by the semi-empirical Simon-Glatzel law \cite{100}, which is 
\begin{gather}
T^m=T^*\left(\frac{P-P^*}{b}+1\right)^{\frac{1}{c}}.
\label{eq:21}
\end{gather}
Our fitted results for $T^*$, $P^*$, $b$ and $c$ are listed in Table \ref{tbl:1}. They can be utilized to consider the thermophysical state of rocky exoplanets (see Appendix C). 

\begin{table}[htp]
  \caption{\ Simon-Glatzel parameters for hcp iron and bcc iron.}
  \label{tbl:1}
  \begin{ruledtabular}
  \begin{tabular}{ccccc}
    Phase & $T^*$ & $P^*$ & $b$ & $c$ \\
    \colrule
    hcp & 1811 K & 0 GPa & 16.88 GPa & 2.44 \\
    bcc & 1811 K & 0 GPa & 16.86 GPa & 2.38 \\
  \end{tabular}
  \end{ruledtabular}
\end{table}

\section{Conclusion}
The SMM-WHEP has been successfully developed to shed light on the melting behaviors of iron under intense compression up to 4000 GPa. This one-phase approach enables us to directly infer high-pressure melting properties from equation-of-state parameters. On that basis, we have quantitatively explained previous DAC/SW experiments, DFT/DMC simulations, and SMM calculations with minimal computational cost. Therefore, our theoretical information about molten iron would be valuable for advancing knowledge of planetary interiors. It is practicable to extend our SMM-WHEP analyses to more complex minerals.

\appendix
\section*{Appendix A: The mass-radius relationship}
\setcounter{equation}{0}
\renewcommand{\theequation}{A\arabic{equation}}
Mass ($M$) and radius ($R$) are the only observable characteristics for most super-Earths. To predict their chemical structures, one needs to compare the measured results for $M$ and $R$ with the calculated $M$-$R$ curves of candidate materials \cite{101}. Notwithstanding, theoretical $M$-$R$ diagrams often show wide variations owing to the lack of EOS data at the TPa regime \cite{102,103,104,105,106,107,108,109,110,111,112}. In this Appendix, we adopt the SMM EOS to determine the $M$-$R$ relationship of homogeneous iron planets. Fundamentally, their hydrostatic equilibrium can be well described by \cite{108}
\begin{eqnarray}
\frac{dP\left(r\right)}{dr}=-g\left(r\right)\rho\left(r\right),
\label{eq:A1}
\end{eqnarray}
where $g$ is the Newtonian gravitational acceleration. Employing the Gauss theorem gives us \cite{108}
\begin{eqnarray}
\frac{dg\left(r\right)}{dr}=4\pi G\rho\left(r\right)-2\frac{g\left(r\right)}{r},
\label{eq:A2}
\end{eqnarray}
where $G$ is the Newtonian constant of gravitation. For a given value of the core pressure $P_c$, we numerically solve Equations (\ref{eq:13}), (\ref{eq:A1}), and (\ref{eq:A2}) from the planetary center ($r\rightarrow0$, $P\rightarrow P_c$, $g\rightarrow0$) to the planetary surface ($r\rightarrow R$, $P\rightarrow0$, $g\rightarrow g_s$). After obtaining $R$ and $g_s$, we can readily compute $M$ by \cite{108}
\begin{eqnarray}
M=\frac{g_s}{G}R^2.
\label{eq:A3}
\end{eqnarray}

\begin{figure}[htp]
\setcounter{figure}{0}
\renewcommand{\thefigure}{A\arabic{figure}}
\includegraphics[width=12 cm]{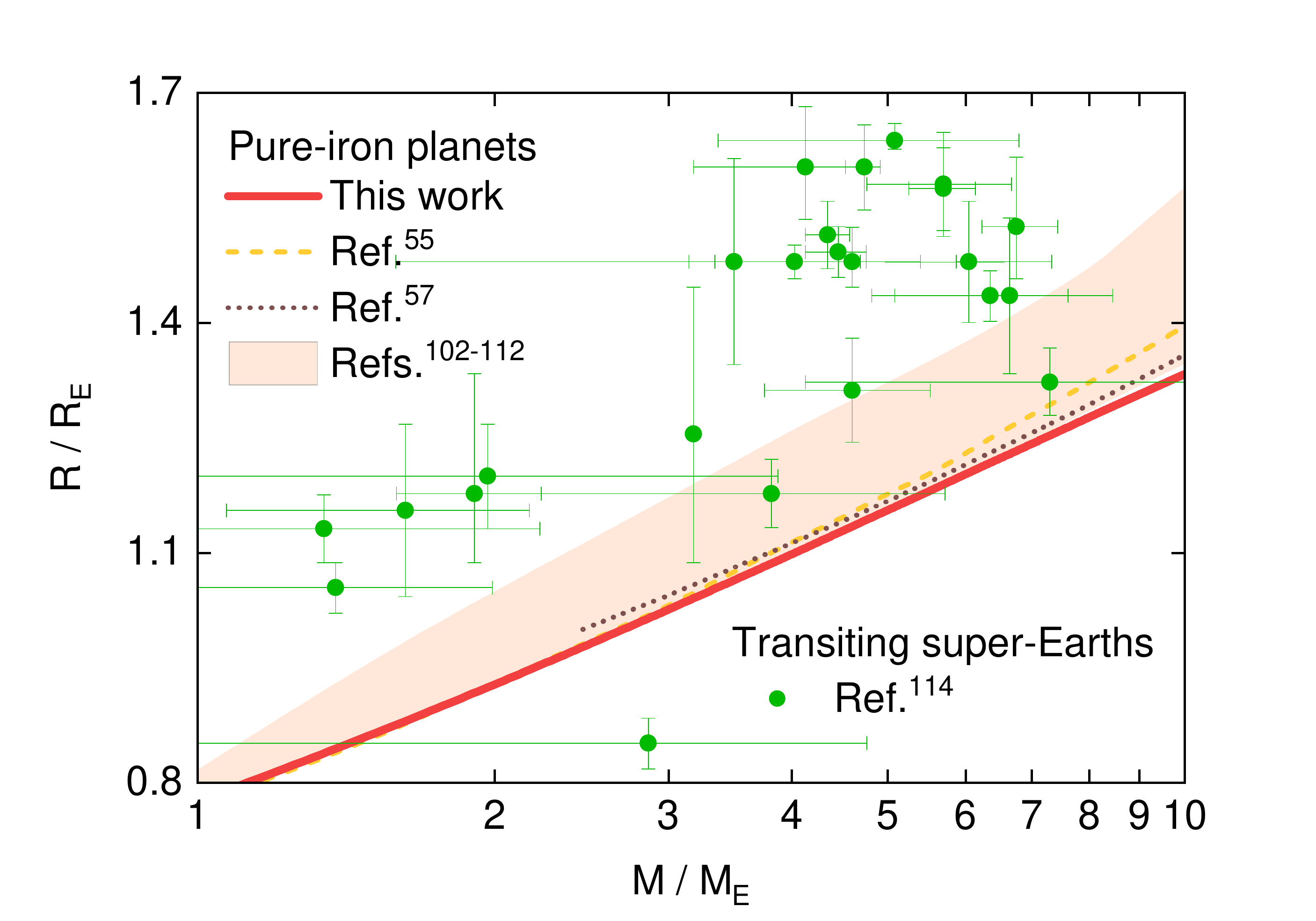}
\caption{\label{fig:A1}(Color online) The mass-radius correlation of hypothetical pure-iron planets provided by our SMM study and Refs.\cite{55,57,102,103,104,105,106,107,108,109,110,111,112}. Experimental data for transiting super-Earths \cite{114} are also plotted for comparison. }
\end{figure}

Figure \ref{fig:A1} illustrates how $R/R_E$ depends on $M/M_E$ in the case of pure-iron celestial bodies. Here, $R_E$ and $M_E$ are the Earth's radius and mass, respectively. It is clear to see that $R/R_E$ grows monotonically with $M/M_E$. For instance, $R/R_E$ increases from 0.77 to 1.33 when $M/M_E$ climbs from 1 to 10. Our SMM analyses are in consonance with the \textit{ab initio} simulations of Hakim \textit{et al.} \cite{57}. The maximum difference between these approaches is merely 2.30 \%.

Notably, some prior studies \cite{113} have suggested the existence of super-dense exoplanets. These anomalous objects are markedly heavier than homogeneous iron spheres of the same size. However, according to our theoretical calculations, the probability of detecting them is extremely low. Most of the observed $M$-$R$ points \cite{114} locate above our $M$-$R$ boundary, except for a few rare circumstances with large uncertainties. Our conclusion is forcefully supported by the state-of-the-art RW experiments of Smith \textit{et al.} \cite{55}.

\appendix
\section*{Appendix B: The age of the Earth's inner core}
\setcounter{equation}{0}
\renewcommand{\theequation}{B\arabic{equation}}
\begin{table}[htp]
\setcounter{table}{0}
\renewcommand{\thetable}{B\arabic{table}}
  \caption{\ Thermophysical parameters needed for the Buffett model \cite{115}. $dT^m/dP$, $dT^s/dP$, $L$, $\gamma_G$, $K_S$, $C_P$, and $\Delta\rho_{ICB}$ are considered at the ICB, whereas $Q$ is defined at the Earth's core-mantle boundary.}
  \label{tbl:B1}
  \begin{ruledtabular}
  \begin{tabular}{cccc}
    Symbol & Physical meaning & Specific value & Reference \\
    \colrule
    $dT^{m}/dP$ & Clapeyron slope & $7.7$ $\mathrm{K}\,\mathrm{GPa}^{-1}$ & This work\\
    $dT^{s}/dP$ & Adiabatic gradient & $5.8$ $\mathrm{K}\,\mathrm{GPa}^{-1}$ & Ref.\cite{4}\\
    $L$ & Fusion enthalpy & $10^6$ $\mathrm{J}\,\mathrm{kg}^{-1}$ & Ref.\cite{85}\\
    $\gamma_G$ & Gruneisen parameter& $1.39$ & Ref.\cite{116}\\
    $K_S$ & Adiabatic bulk modulus& $1306$ GPa & Ref.\cite{116}\\
    $C_P$ & Isobaric heat capacity & $790$ $\mathrm{J}\,\mathrm{kg}^{-1}\,\mathrm{K}^{-1}$ & Ref.\cite{116}\\
    $r_{IC}$ & Inner-core radius & $1221$ km & Ref.\cite{117} \\
    $r_{OC}$ & Outer-core radius & $3480$ km & Ref.\cite{117}\\
    $\rho_{avg}$ & Outer-core mean density& $12000$ $\mathrm{kg}\,\mathrm{m}^{-3}$ & Ref.\cite{117}\\
    $\Delta\rho_{ICB}$ & Compositional density jump& $620$ $\mathrm{kg}\,\mathrm{m}^{-3}$ & Ref.\cite{118}\\
    $Q$ & Heat flow & $10$ TW & Ref.\cite{119} \\
  \end{tabular}
  \end{ruledtabular}
\end{table}

The Earth's core has been solidified from the bottom up. To estimate the interval $\tau$ between the crystallization onset and the present time, we use the energy-conservation model of Buffett, which is \cite{115}
\begin{eqnarray}
\tau=\frac{\mathcal{M}}{Q}\left[\left(\frac{r_{IC}}{r_{OC}}\right)^2+\mathcal{L}\left(\frac{r_{IC}}{r_{OC}}\right)^3+\mathcal{G}\left(\frac{r_{IC}}{r_{OC}}\right)^3\right],
\label{eq:B1}
\end{eqnarray}
where
\begin{eqnarray}
\mathcal{M}=4\pi\left(\frac{1}{3}-\frac{1}{5}\frac{Ar_{OC}^2\gamma_G}{K_S}\right)\rho_{avg}C_PAr_{OC}^5\left(\frac{dT^m}{dP}-\frac{dT^s}{dP}\right),\nonumber\\
A=\frac{2\pi}{3}G\rho_{avg}^2,\quad\mathcal{L}=\frac{A\pi}{3}\frac{\rho_{avg}Lr_{OC}^3}{\mathcal{M}},\quad\mathcal{G}=\frac{4\pi}{5}\frac{Ar_{OC}^5}{\mathcal{M}}\frac{\Delta\rho_{ICB}}{\rho_{avg}}.
\label{eq:B2}
\end{eqnarray}
Details about these thermophysical parameters are presented in Table \ref{tbl:B1} \cite{4,85,116,117,118,119}. On that basis, we get $\tau=0.44$ Gyr. This figure implies that our planet has a young solid inner core \cite{80}. Our physical picture is very consistent with recent DFT and paleomagnetic explorations \cite{120,121} ($\tau=0.4-0.7$ Gyr).  

\appendix
\section*{Appendix C: The state of terrestrial exoplanets}
\setcounter{equation}{0}
\renewcommand{\theequation}{C\arabic{equation}}
It is well-known that the habitability of super-Earths depends crucially on the thermodynamic state of their cores \cite{122}. Unfortunately, too many contradictory scenarios have been proposed. For example, Valencia \textit{et al.} \cite{123} have predicted that super-Earth cores would be completely solid. In contrast, Sotin \textit{et al.} \cite{124} have argued that these systems would be entirely liquid. Additionally, some other scientists have suggested that the core solidification would occur from the top down via iron snowflake-like mechanisms \cite{125}. These persisting controversies principally stem from the scarcity of ultrahigh-pressure melting information. 

Here, we apply SMM-WHEP results to deal with the mentioned issues. First of all, we evaluate the depression effect of light elements (e.g., S, O, and Si) on the melting transition of hcp iron via the cryoscopic equation as \cite{126}
\begin{eqnarray}
T^m_{alloy}=\frac{T^m}{1-\ln\left(1-x^*\right)}=\frac{T^*}{1-\ln\left(1-x^*\right)}\left(\frac{P-P^*}{b}+1\right)^{\frac{1}{c}},
\label{eq:C1}
\end{eqnarray}
where $x^*=0.13$ is the mole fraction of impurities. Next, the temperature distribution in deep-planetary interiors is determined by the adiabatic method as \cite{126}
\begin{eqnarray}
\frac{\partial\ln T_{core}}{\partial\ln\rho_{core}}=\gamma_G.
\label{eq:C2}
\end{eqnarray}
The core density $\rho_{core}$ and the Gruneisen parameter $\gamma_G$ are taken from high-quality \textit{ab initio} datasets of Stixrude \cite{127}. Meanwhile, the initial value of $T_{core}$ at the core-mantle interface is deduced from the latest SW measurements of Fei \textit{et al.} \cite{128}. Finally, the fundamental physical structure of super-Earth cores is elucidated by comparing $T^m_{alloy}$ with $T_{core}$.

Figure \ref{fig:C1} shows our representative numerical calculations for a ten-Earth-mass celestial body with a partially molten silicate mantle. It is conspicuous that $T^m_{alloy}$ is lower than $T_{core}$ at the uppermost core ($P\sim1350$ GPa \cite{18}). Nevertheless, the melting slope is appreciably higher than the adiabatic gradient. These events lead to the intersection of the melting curve and the adiabatic line at around 2500 GPa. In other words, similar to our planet, the studied super-Earth has both a molten outer core and a frozen inner core. Thus, the generation of its magnetic field can be efficiently promoted by the churning of liquid iron-based alloys \cite{129,130}. This magnetic shield helps protect organic life forms from dangerous cosmic radiation \cite{129,130}. Our findings concur with the leading-edge SW experiments of Kraus \textit{et al.} \cite{40}.

\begin{figure}[htp]
\setcounter{figure}{0}
\renewcommand{\thefigure}{C\arabic{figure}}
\includegraphics[width=12 cm]{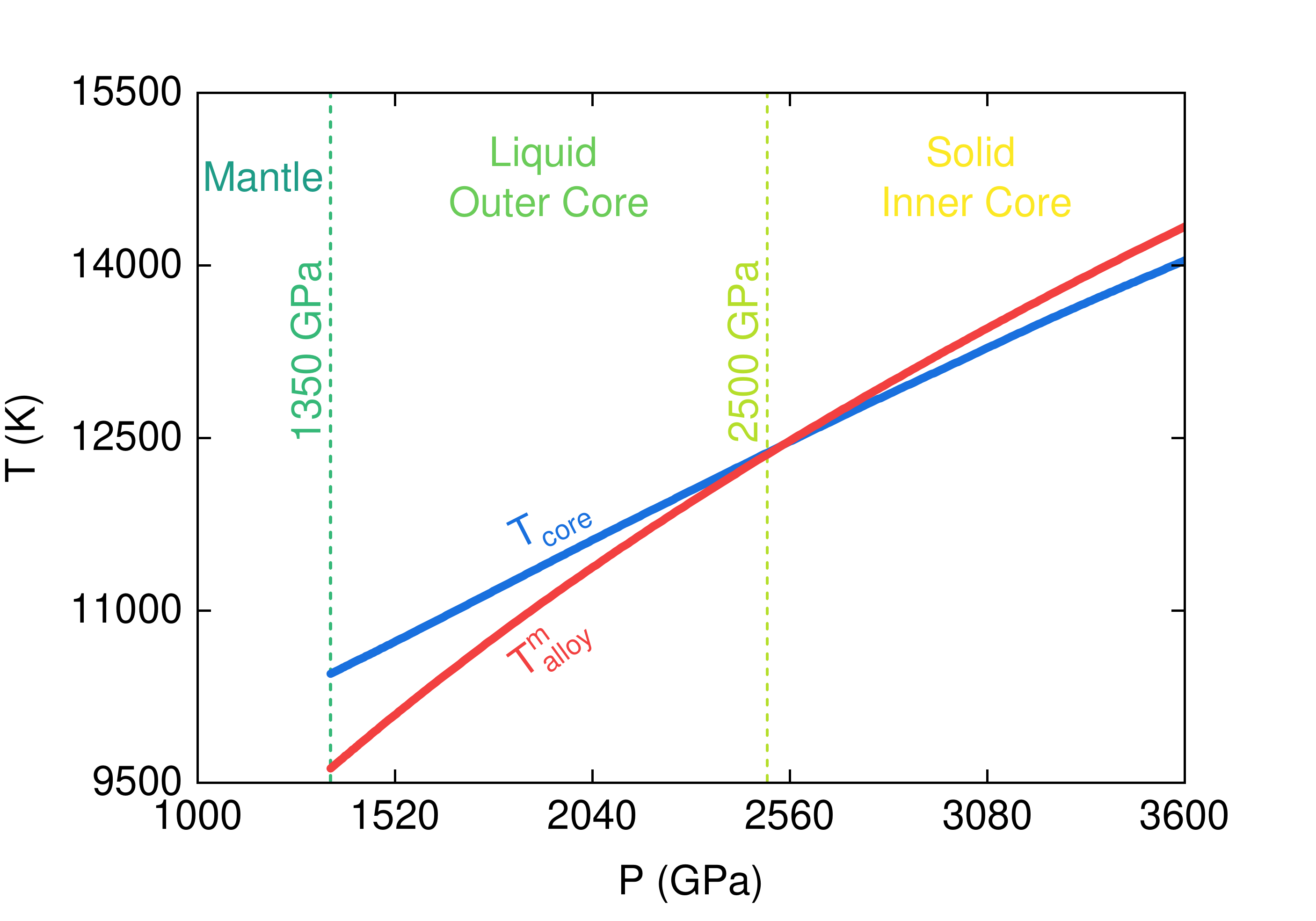}
\caption{\label{fig:C1}(Color online) Our theoretical predictions for the core crystallization of a telluric exoplanet with $M=10M_E$.}
\end{figure}

\end{document}


\title{Electronic Supplementary Information for \textit{Theoretical predictions of melting behaviors of hcp iron up to 4000 GPa}}
\author{Tran Dinh Cuong}
\email{cuong.trandinh@phenikaa-uni.edu.vn}
\affiliation{Faculty of Materials Science and Engineering, Phenikaa University, Hanoi 12116, Vietnam.}
\author{Nguyen Quang Hoc}
\affiliation{Faculty of Physics, Hanoi National University of Education, Hanoi 11310, Vietnam.}%
\author{Nguyen Duc Trung}
\affiliation{Faculty of Physics, Hanoi National University of Education, Hanoi 11310, Vietnam.}%
\author{Nguyen Thi Thao}
\affiliation{Faculty of Physics, Hanoi National University of Education, Hanoi 11310, Vietnam.}%
\author{Anh D. Phan}
\affiliation{Faculty of Materials Science and Engineering, Phenikaa University, Hanoi 12116, Vietnam.}
\affiliation{Phenikaa Institute for Advanced Study (PIAS), Phenikaa University, Hanoi 12116, Vietnam.}
\maketitle

\section{The effectiveness of the SMM-WHEP SCHEME}
Recently, physicists have succeeded in constraining the melting curve of aluminum, copper, and vanadium by different methods. Hence, in the present Supplemental Material, these metallic systems are utilized to prove the quality of the SMM-WHEP scheme.

\begin{table}[htp]
\renewcommand{\thetable}{S\arabic{table}}
  \caption{\ The Morse parameters for aluminum, copper, and vanadium [S1, S2].}
  \label{tbl:S1}
  \begin{ruledtabular}
  \begin{tabular}{ccccc}
    Metal & Structure & $D$ (eV) & $\alpha$ (\AA$^{-1}$) & $r_0$ (\AA) \\
    \colrule
    Al & fcc & 0.2703  & 1.1646  & 3.253  \\
    Cu & fcc & 0.3429  & 1.3588 & 2.866 \\
    V & bcc & 0.5871 & 1.1912  & 2.873  \\
  \end{tabular}
  \end{ruledtabular}
\end{table}

First, we extract the Morse parameters from Refs.[S1, S2] to calculate the SMM EOS (see Table S1). Figure S1 shows that the Morse potential can help us capture the atomic rearrangement of aluminum, copper, and vanadium with high accuracy [S3$-$S5]. Based on the Vinet fitting [S6], we acquire the isothermal bulk modulus $K_0$ and its pressure derivative $K_0'$ in Table S2.

\begin{table}[htp]
\renewcommand{\thetable}{S\arabic{table}}
  \caption{\ The input data for WHEP calculations in the case of aluminum, copper, and vanadium. The detailed information about $(P_1,T_1)$ and $(P_2,T_2)$ can be found in Refs.[S7, S8].}
  \label{tbl:S2}
  \begin{ruledtabular}
  \begin{tabular}{cccccccc}
    Metal & Structure & $K_0$ (GPa) & $K_0'$ & $P_1$ (GPa) & $T_1$ (K) & $P_2$ (GPa) & $T_2$ (K) \\
    \colrule
    Al & fcc & 64.97  & 5.12  & 0 & 933 & 0.75 & 995  \\
    Cu & fcc & 125.13  & 5.10 & 0 & 1356 & 0.75 & 1395 \\
    V & bcc & 136.08 & 4.61  & 0 & 2183 & 40 & 3200  \\
  \end{tabular}
  \end{ruledtabular}
\end{table}

Next, we enter $K_0$ and $K_0'$ into the WHEP model to locate the melting point of aluminum, copper, and vanadium between 0 and 500 GPa. As presented in Figures S2, S3, and S4, this strategy enables us to redescribe all previous experimental and computational results [S7$-$S22] at a quantitative level. Besides, it is explicitly observed that our modified WHEP version (Equation (17)) is superior to the original one (Equation (16)). The melting-temperature underestimation is adequately addressed by replacing the Murnaghan EOS with the SMM EOS. 

The aforementioned evidence verifies the reliability and flexibility of SMM-WHEP analyses in investigating the melting process at intense pressures.

\section{Two-phase simulations for iron}
To strengthen SMM-WHEP predictions for hcp-iron, we perform classical molecular dynamics (CMD) simulations via the open-source LAMMPS package [S23]. The two-phase approach (TPA) [S24$-$S28] is adopted to determine the melting temperature of hcp-iron with the DFT-based Morse pairwise potential [S2] ($D=0.6317$ eV, $\alpha=1.4107$ \AA$^{-1}$, and $r_0=2.6141$ \AA\/). Physically, this modern method allows us to overcome the overheating problem of conventional one-phase calculations by adding surface effects [S24$-$S28]. Moreover, it is widely demonstrated that the CMD-TPA can give us exact melting information about the chosen interatomic potential [S24$-$S28].

Technical details about the CMD-TPA are summarized in Figure S5. Overall, our computational process at a given pressure can be divided into two principal stages: (i) preparation (Figure S5(a)) and (ii) production (Figure S5(b)). In stage (i), we construct two identical perfect hcp supercells, each containing 2000 atoms. Then, one of these lattices is isochorically heated to an extreme temperature to melt completely. Finally, we put the obtained solid and liquid structures together in our simulated box. A small gap is created between them to prevent atoms from overlapping and lessen the strain at the common interface. In stage (ii), our two-phase system is simulated in the $NPT$ ensemble. Each CMD run is done within 100 ps (100000 timesteps) to find the equilibrium volume. On that basis, the melting transition is detected via an abrupt change in the volume-temperature diagram. Note that the TPA is quite different from the coexistence method [S29], where the solid and liquid parts can coexist under the $NVE$ condition. For the TPA, the system is equilibrated towards a monophase. To be more specific, it is solidified at $T<T^m$ and liquefied at $T>T^m$. 

Our CMD-TPA results are compared to SMM-WHEP data in Figure S6. One can readily see an excellent agreement between the above approaches under Earth-core conditions. For instance, at the ICB, CMD-TPA simulations yield $T^m=6050$ K, which is merely lower than the SMM-WHEP output by 3.09 \%. This quantitative consistency reaffirms the ``steep-melting-line" scenario for iron as well as the accuracy of SMM-WHEP approximations.



\begin{figure}[htp]
\renewcommand{\thefigure}{S\arabic{figure}}
\includegraphics[width=15 cm]{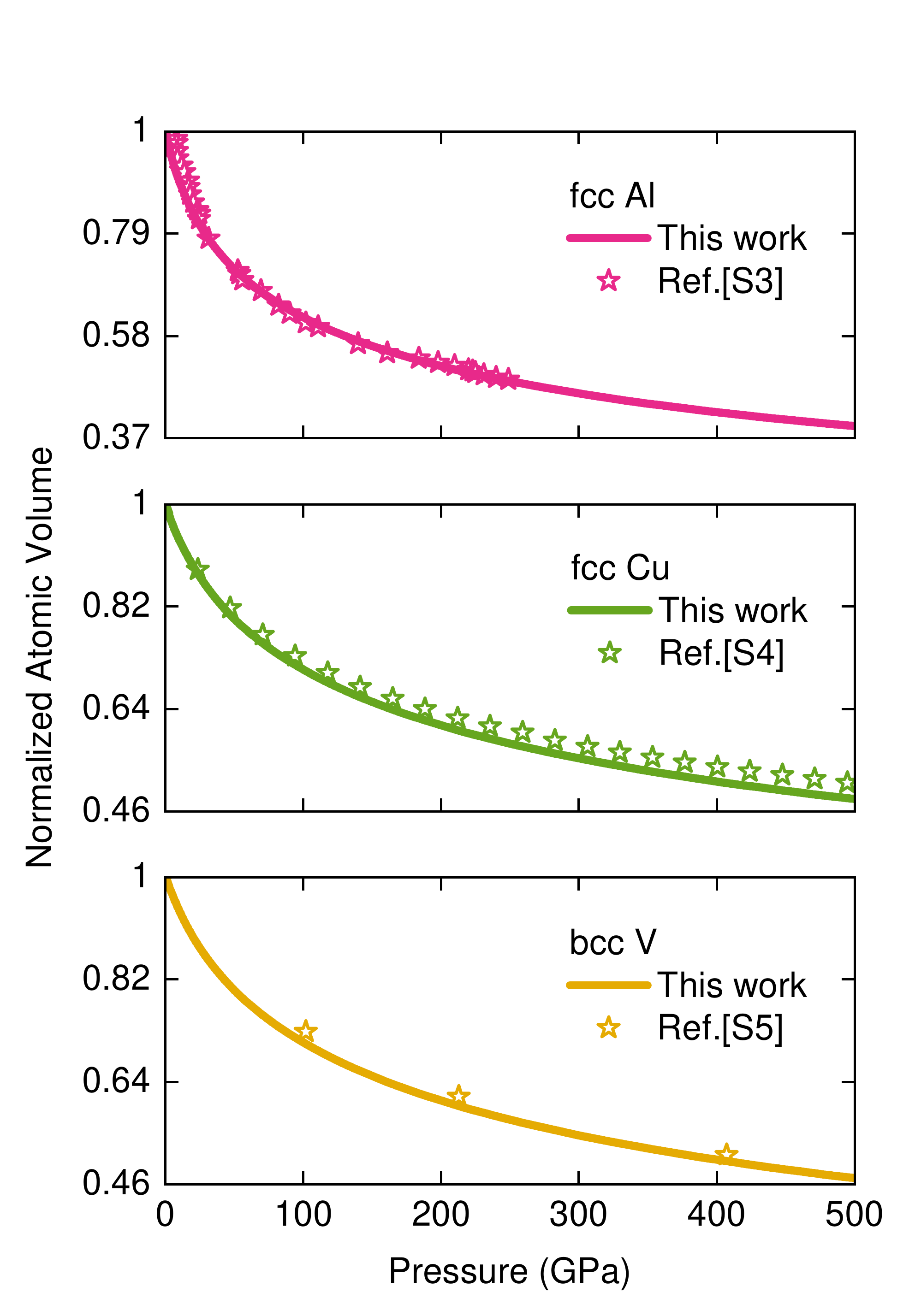}
\caption{(Color online) The EOS of aluminum, copper, and vanadium given by our SMM calculations, DAC experiments [S3], RW measurements [S4], and DFT simulations [S5].}
\end{figure}

\begin{figure}[htp]
\renewcommand{\thefigure}{S\arabic{figure}}
\includegraphics[width=15 cm]{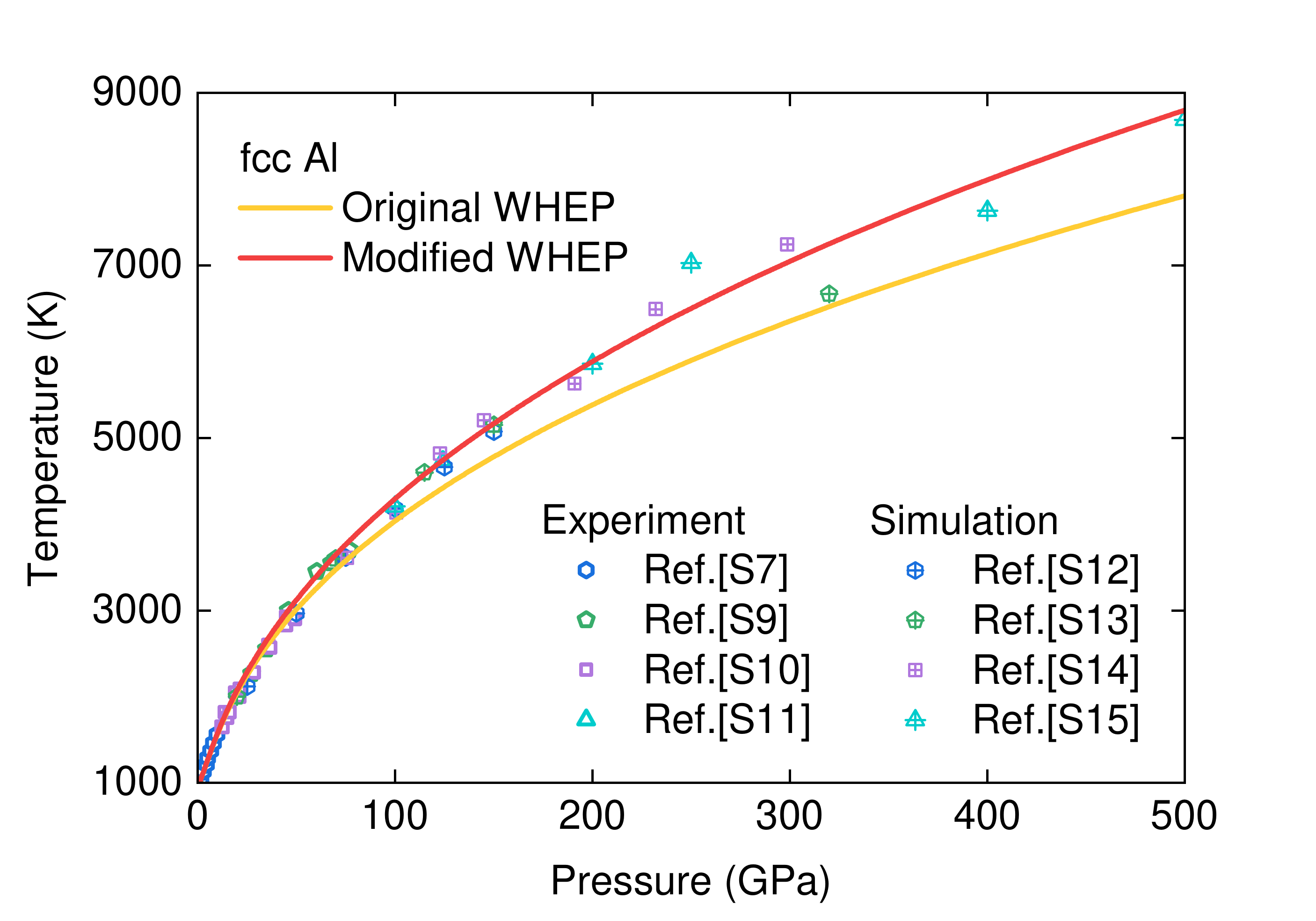}
\caption{(Color online) The melting diagram of aluminum obtained from our SMM-WHEP theory and other approaches [S7, S9$-$S15].}
\end{figure}

\begin{figure}[htp]
\renewcommand{\thefigure}{S\arabic{figure}}
\includegraphics[width=15 cm]{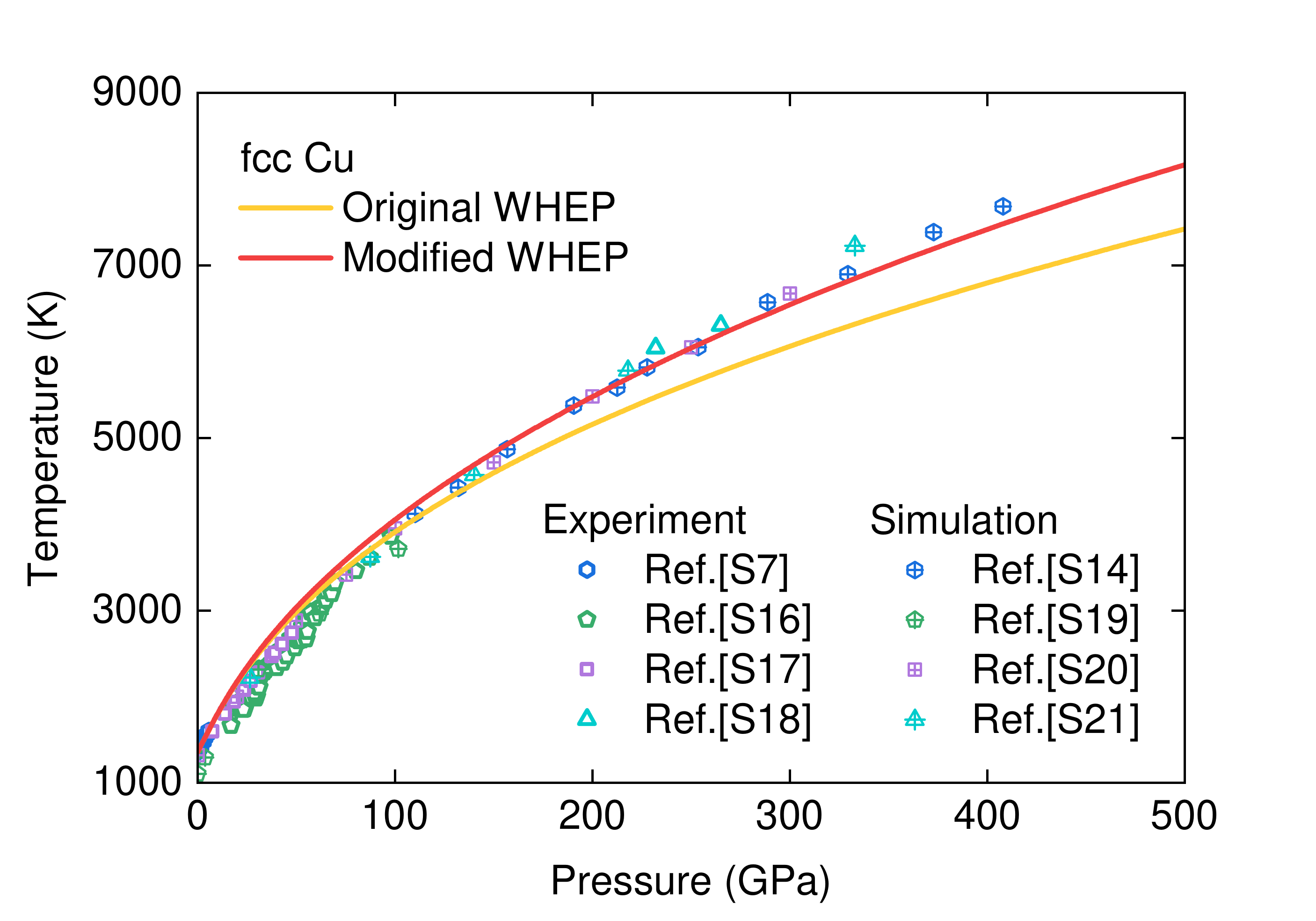}
\caption{(Color online) Our SMM-WHEP outputs and the prior measured/simulated results [S7, S14, S16$-$S21] for the melting profile of copper.}
\end{figure}

\begin{figure}[htp]
\renewcommand{\thefigure}{S\arabic{figure}}
\includegraphics[width=15 cm]{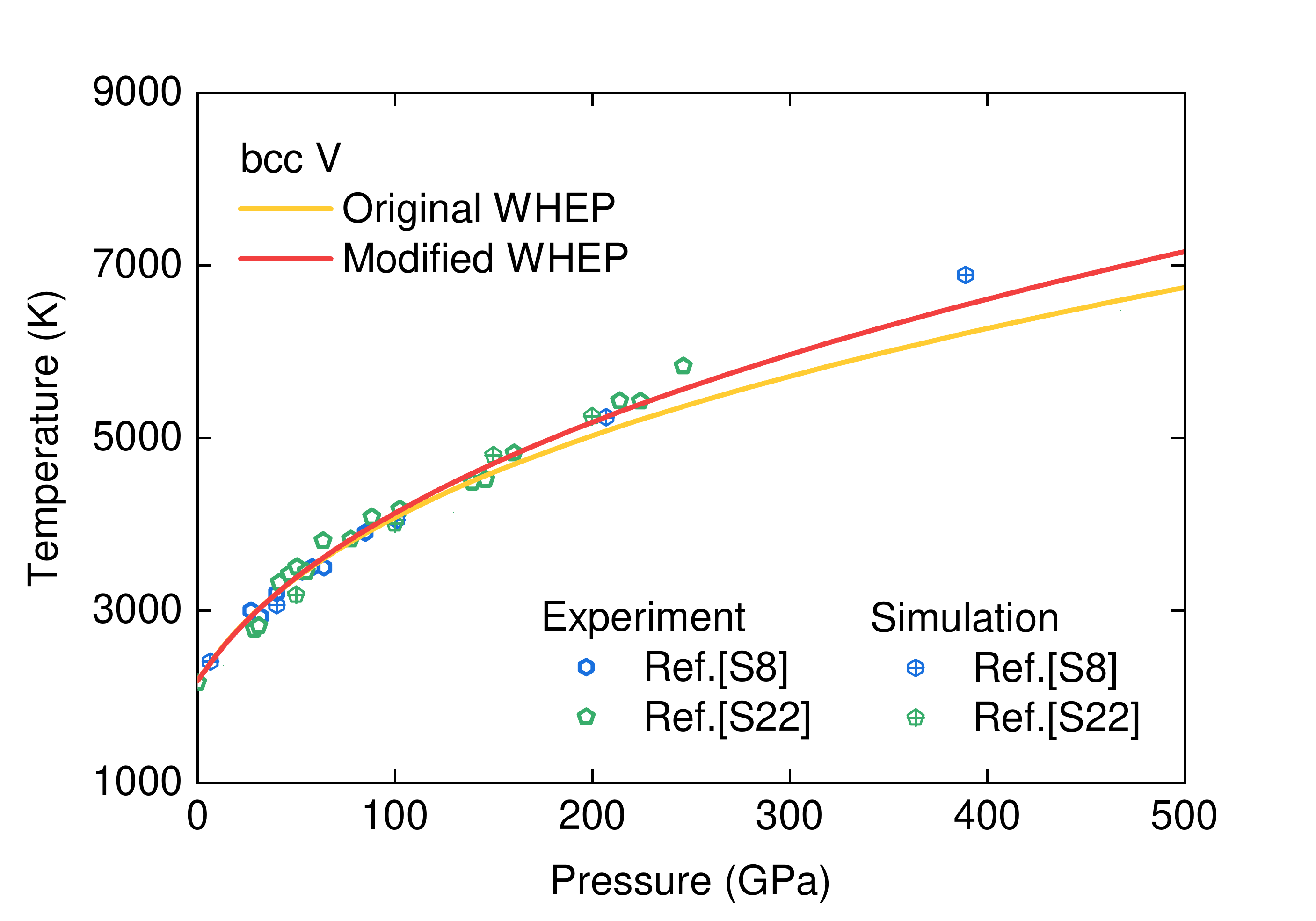}
\caption{(Color online) Compression effects on the melting temperature of vanadium provided by our SMM-WHEP analyses and recent studies [S8, S22].}
\end{figure}

\begin{figure}[htp]
\renewcommand{\thefigure}{S\arabic{figure}}
\includegraphics[width=15 cm]{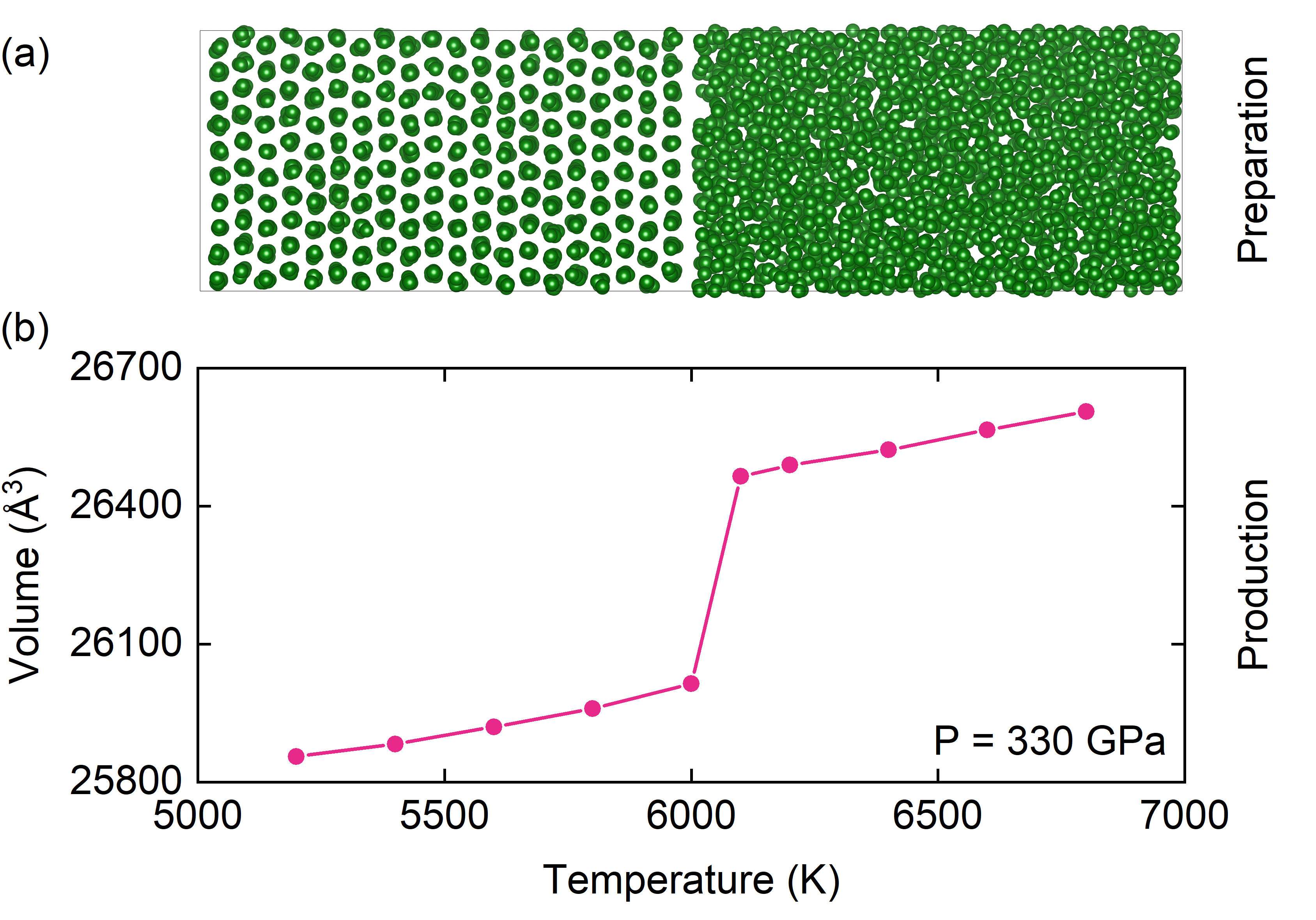}
\caption{(Color online) Main stages to derive the high-pressure melting point of hcp-iron from our CMD-TPA simulations. Structural visualization is made by VESTA [S30].}
\end{figure}

\begin{figure}[htp]
\renewcommand{\thefigure}{S\arabic{figure}}
\includegraphics[width=15 cm]{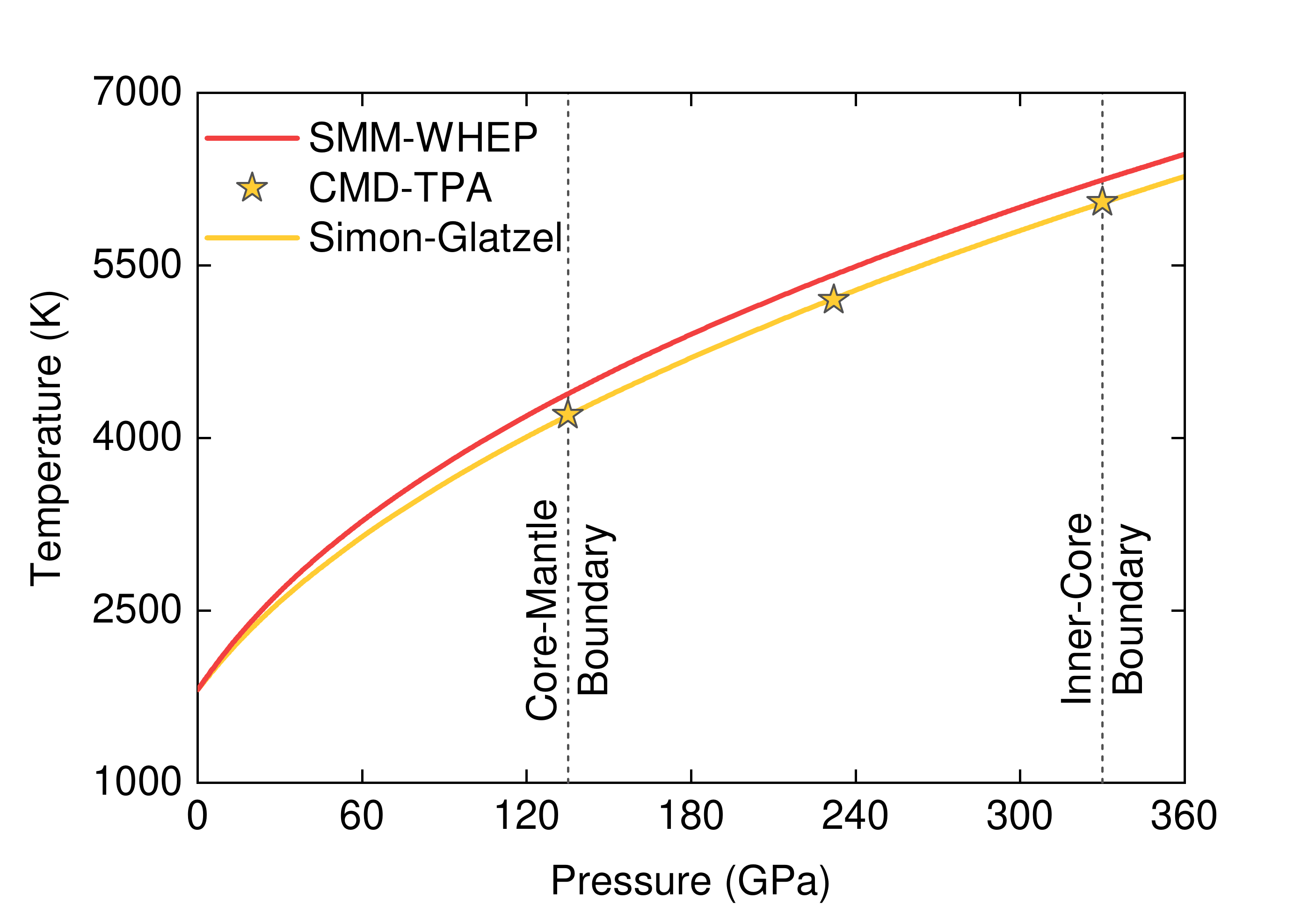}
\caption{(Color online) Comparison between CMD-TPA and SMM-WHEP results for the melting curve of hcp-iron. Our CMD-TPA data can be well described by the Simon-Glatzel equation as $T^m(K)=1811\left(\cfrac{P}{26.02}+1\right)^{0.46}$, where $P$ is in GPa.}
\end{figure}